\documentclass[twocolumn,epjc3]{svjour3}
\usepackage{amsmath}
\usepackage{amsfonts}
\usepackage{multirow}
\usepackage{enumerate}
\usepackage{color}
\usepackage{graphicx}
\usepackage{epstopdf}
\usepackage{dcolumn}
\usepackage{bm}
\usepackage{amssymb}
\usepackage{hyperref}
\usepackage{cite}
\usepackage{url}
\hypersetup{unicode,bookmarksnumbered=true,colorlinks,
	linkcolor=black,%
	anchorcolor=red,
	urlcolor=black,
	citecolor=blue,
}
\usepackage{subfigure}

\begin{document}
\newcommand{\red}[1]{\textcolor{red}{#1}}
\newcommand{\green}[1]{\textcolor{green}{#1}}
\newcommand{\blue}[1]{\textcolor{blue}{#1}}
\newcommand{\cyan}[1]{\textcolor{cyan}{#1}}
\newcommand{\purple}[1]{\textcolor{purple}{#1}}
\newcommand{\yellowbox}[1]{\colorbox{yellow}{#1}}
\newcommand{\purplebox}[1]{\colorbox{purple}{#1}}
\newcommand{\yellow}[1]{\textcolor{yellow!70!red}{#1}}
\title{Diagnostics for generalized power-law torsion-matter coupling $f(T)$ model
}
\author{Xiang-Hua Zhai\thanksref{e1,addr} \and Qiang Wen\thanksref{addr} \and Rui-Hui Lin\thanksref{e2,addr} \and Xin-Zhou Li\thanksref{e3,addr}}
\thankstext{e1}{e-mail: zhaixh@shnu.edu.cn}
\thankstext{e2}{e-mail: linrh@shnu.edu.cn}
\thankstext{e3}{e-mail: kychz@shnu.edu.cn}
\institute{Shanghai United Center for Astrophysics (SUCA), Shanghai Normal University,
	100 Guilin Road, Shanghai 200234, China\label{addr}}
\maketitle

\begin{abstract}
The currently accelerated expansion of our Universe is unarguably
one of the most intriguing problems in today's physics research.
Two realistic non-minimal torsion-matter
coupling $f(T)$ models have been established
and studied in our previous papers [Phys. Rev. D92, 104038(2015) and Eur. Phys. J. C77, 504(2017)]
aiming to explain this ``dark energy'' problem.
In this paper,
we study the generalized power-law
torsion-matter coupling $f(T)$ model.
Dynamical system analysis shows that
the three expansion phases of the Universe,
i.e. the radiation dominated era,
the matter dominated era and
the dark energy dominated era,
can all be reproduced in this generalized model. By using the statefinder and $Om$ diagnostics, we find that the different cases of the model can be distinguished from each other and from  other dark energy models
such as the two models in our previous papers, $\Lambda$CDM, quintessence and Chaplygin gas. Furthermore, the analyses also show that all kinds of generalized power-law torsion-matter coupling model are able to cross the $w=-1$  divide from below to above, thus the decrease of the energy density resulting from the crossing of $w$ will make the catastrophic fate of the Universe avoided and a de Sitter expansion fate in the future will be approached.

\end{abstract}


\section{Introduction}\label{secI}
It is suggested by observations
including the type Ia supernovae (SNIa), cosmic microwave background radiation (CMB), baryon acoustic oscillations (BAO) and large-scale structure (LSS) that our Universe is now in an acceleration phase of expansion.
And this acceleration is no doubt one of the most intriguing problems
in current physics research.
To explain this weird phenomenon, physicists introduced the cosmological constant $\Lambda$ back again, and constructed a new standard model ($\Lambda$CDM), also known as the concordance model. However, the value of $\Lambda$ is too small to be explained by any current fundamental theories. To alleviate this troublesome problem, various dynamical dark energy theories have been proposed, in which the energy composition is dependent on time. But these exotic fields are still phenomenological, lacking theoretical foundations. Besides adding unknown fields, there is another kind of theories known as modified gravity, such as
$f(R)$ gravities\cite{Sotiriou:2008rp,DeFelice:2010aj},
the modified Newtonian dynamics (MOND) cosmology\cite{Zhang:2011uf},
Poincar\'e gauge theory \cite{Li:2009zzc,Ao:2010mg,Ao:2011kc}.

On the other hand,
the Teleparallel Equivalent of General Relativity(TEGR) \cite{Einstein:1928}
was constructed by Einstein.
To modify TEGR,
the simplest scheme is the $f(T)$ gravity
\cite{Ferraro2006,Linder2010,Aly:2015fda,Cai2015,Farrugia:2016,Nunes:2016plz,Oikonomou:2016jjh,Lin2016,Hohmann2017,Qi:2017xzl,Sk:2017ucb,Capozziello:2017bxm,Oikonomou:2017isf},
which extends the torsion scalar $T$
in the Lagrangian to an arbitrary function
$f(T)$.
Although TEGR is equivalent to GR,
modified gravities based on $T$
are not similar to $f(R)$ in many aspects,
among which there is an important advantage
that the equations of motion
are second order instead of fourth order.

Under the framework of Teleparallelism,
further extensions to $f(T)$ gravities have also been proposed,
such as $f(T,\mathcal T)$ gravities
\cite{Pace:2017aon,Zhai:2017yqd,Harko:2014aja},
where $\mathcal T$ is the trace of the matter energy-momentum tensor $\mathcal{T}_{\mu\nu}$, and
$f(T,T_G)$ theories
\cite{Kofinas:2014owa,Bahamonde2016,Capozziello2016,Jawad2015,Kofinas:2014daa},
where $T_G$ is the teleparallel equivalence of the Gauss-Bonnet term. One of these extensions is the non-minimal torsion-matter coupling $f(T)$ gravity
\cite{Harko2014,Carloni2015,Feng2015,Lin2017}, in which we investigated the evolution of the Universe \cite{Feng2015} and the Solar system tests \cite{Lin2017} for realistic models.

In this paper,
we generalize the models in Refs.\cite{Feng2015,Lin2017} to one including multiple terms of power-law of the torsion scalar $T$.
We establish the dynamical systems for
several cases of the model,
and analyze the critical points and the phase space.
The different phases of the Universe evolution
are reproduced and confirmed. Then we compare them with
the two models in Refs.\cite{Feng2015,Lin2017},
$\Lambda$CDM, quintessence and Chaplygin gas models
using statefinder and $Om$ diagnostics.
We find that the different cases of our model be distinguished from each other and
from other models. Furthermore,we find that the evolution of equation of state parameter $w$ with cosmic time can cross the divide $w=-1$ from below to above and approach to $w=-1$ in the future. In other words, all kinds of generalized power-law torsion-matter coupling $f(T)$ model are able to avoid the catastrophic big or little rip\cite{Xi2012}.

The main contents of this paper are as follows.
In Section \ref{secII},
we briefly review the non-minimal torsion-matter coupling
$f(T)$ gravity.
The model and its cosmological fit
are given in Sec. \ref{secIII}.
In Sec. \ref{secIV},
we study our model using dynamical system analysis.
We use statefinder and $Om$ diagnostics to
investigate our model in Sects. \ref{secV} and \ref{secVI}, respectively.
Finally, Sec. \ref{secVII} contains our concluding remarks.

\section{Equations of motion}\label{secII}
In this section, we make a brief review of the equations of motion for the non-minimal torsion-matter coupling $f(T)$ gravity. In order to describe the evolution history of the Universe including the era before matter-radiation equality, we adopt the action provided in Ref. \cite{Feng2015}:
\begin{eqnarray}
\begin{split}
S=\frac{1}{2\kappa}\int   |e|&\{[1+f_{1}(T)]T+[1+f_{2}(T)]\mathcal{L}_{M}\\&+\mathcal{L}_{r}\}d^{4}x,\label{S}
\end{split}
\end{eqnarray}
where $\kappa=8\pi G$, $|e|=det(e^{A}_{\mu})$, $f_{1}(T)$ and $f_{2}(T)$ are arbitrary functions of the torsion scalar $T$, and $\mathcal{L}_{M}$ and $\mathcal{L}_{r}$ are the matter and radiation Lagrangian densities. Here the inverse vierbeins $e^{A}_{\mu}$ satisfy
\begin{eqnarray}
e^{A}_{\ \ \mu}e_{B}^{\ \ \mu}=\delta_{B}^{A},\quad
e_{A}^{\ \ \mu}e^{A}_{\ \ \nu}=\delta_{\nu}^{\mu},
\end{eqnarray}
and they can be used to obtain the metric as
\begin{equation}
g_{\mu\nu}=\eta_{AB}e^{A}_{\ \ \mu}e^{B}_{\ \ \nu},\quad
\eta_{AB}=g_{\mu\nu}e_{A}^{\ \ \mu}e_{B}^{\ \ \nu}.
\end{equation}
Furthermore, the torsion scalar $T$ is defined as
\begin{eqnarray}
T\equiv T^{\alpha}_{\ \ \mu\nu} S_{\alpha}^{\ \ \mu\nu},
\end{eqnarray}
where
\begin{equation}
T^{\alpha}_{\ \ \mu\nu}\equiv e_{A}^{\ \ \alpha}(\partial_{\mu} e^{A}_{\ \ \nu}-\partial_{\nu} e^{A}_{\ \ \mu}),
\end{equation}
and
\begin{equation}
S_{\alpha}^{\ \ \mu\nu}\equiv \frac{1}{2}(K^{\mu\nu}_{\quad \alpha}+\delta^{\mu}_{\alpha}T^{\beta\nu}_{\quad\beta}-\delta^{\nu}_{\alpha}T^{\beta\mu}_{\quad\beta}),
\end{equation}
with
\begin{equation}
K^{\mu\nu}_{\quad \alpha}\equiv -\frac{1}{2}(T^{\mu\nu}_{\quad\alpha}-T^{\nu\mu}_{\quad\alpha}-T_{\alpha}^{\ \ \mu\nu}).
\end{equation}

By varying the action \eqref{S} with respect to the vierbein field and using the unit $\kappa=1$, one can obtain the equation of motion as
\begin{eqnarray}
\begin{split}
\frac{1}{2}&e_{A}^{\ \ \rho}T(1+f_{1})+2(1+f_{1}+Tf^{'}_{1}+f^{'}_{2}\mathcal{L}_{M})\\
&\times[|e|^{-1}\partial_{\mu}(|e|e_{A}^{\ \ \alpha}S_{\alpha}^{\ \ \rho\mu})+e_{A}^{\alpha}T^{\mu}_{\ \ \nu\alpha}S_{\mu}^{\ \ \rho\nu}]\\
&+8(2f^{'}_{1}+Tf_{1}^{''}+f_{2}^{''}\mathcal{L}_{M})(\partial_{\mu}T)e_{A}^{\ \ \alpha}S_{\alpha}^{\ \ \rho\mu}\\
&+8f_{2}^{'}e_{A}^{\ \ \alpha}S_{\alpha}^{\ \ \rho\mu}(\partial_{\mu}\mathcal{L}_{M})\\
=&[(1+f_{2})\mathcal{T}^{\rho}_{\alpha (M)}+\mathcal{T}^{\rho}_{\alpha (r)}]e_{A}^{\ \ \alpha},
\end{split}
\end{eqnarray}
where $\mathcal{T}^{\rho}_{\alpha (M)}$ and $\mathcal{T}^{\rho}_{\alpha (r)}$ are the matter and radiation energy-momentum tensor, respectively, and the prime indicates a derivative with respect to the torsion scalar $T$.

We consider the Friedmann-Robertson-Walker(FRW) Universe in Cartesian coordinates as
\begin{equation}
ds^{2}=-dt^{2}+a(t)^{2}(dx^{i})^{2},
\end{equation}
where $a(t)$ is the scale factor and $i=1,2,3$. A viable choice of tetrad is  $e^{A}_{\ \ \mu}=diag(1,a,a,a)$ \cite{Tamanini:2012hg,Feng2015}. Hence, the torsion scalar can be calculated as $T=-6H^{2}$, where $H=\dot{a}/a$ is the Hubble parameter and overdot denotes the derivative with respective to cosmic time $t$. Choosing the matter Lagrangian density as $\mathcal{L}_{M}=-2\rho_{M}$ \cite{Feng2015,Harko2014,Hervik,Bertolami2008} and assuming that the matter and radiation contents of the Universe are perfect fluids with the energy momentum tensor $\mathcal{T}_{\mu\nu}=pg_{\mu\nu}+(\rho+p)u_{\mu}u_{\nu}$, we can find the equations of motion as
\begin{eqnarray}
\begin{split}
H^{2}=&\frac{1}{3}(1+f_{2})\rho_{M}+\frac{1}{3}\rho_{r}+4H^{2}f_{2}^{'}\rho_{M}\\
&-\frac{1}{6}[Tf_{1}+12H^{2}(f_{1}+Tf_{1}^{'})]\label{Feq1},
\end{split}
\end{eqnarray}
\begin{eqnarray}
\begin{split}
\dot{H}=&-[\frac{1}{2}(1+f_{2})(\rho_{M}+p_{M})+\frac{1}{2}(\rho_{r}+p_{r})\\
&-2f_{2}^{'}H\dot{\rho}_{M}]/[1+f_{1}+Tf_{1}^{'}-2f_{2}^{'}\rho_{M}\\
&-12H^{2}(2f_{1}^{'}+Tf_{1}^{''})+24H^{2}f_{2}^{''}\rho_{M}]\label{Feq2},
\end{split}
\end{eqnarray}
where $\rho_{M}$, $p_{M}$, $\rho_{r}$ and $p_{r}$ are the densities and pressures of matter and radiation. Here $\rho_{M}=\rho_{m}+\rho_{b}$, consisting of cold dark and baryon matters. The continuity equations for matter and radiation
\begin{equation}
\dot{\rho}_{M}+3H\rho_{M}=0,\label{coneq1}
\end{equation}
and
\begin{equation}
\dot{\rho}_{r}+4H\rho_{r}=0\label{coneq2}
\end{equation}
 can be obtained from Eqs. \eqref{Feq1} and \eqref{Feq2}, independent of $f_{1}$ and $f_{2}$.

\section{The model with the best-fit parameters}\label{secIII}
In principle, we can consider the generalization of power-law model as $f_1, f_2 \propto \sum_{n=1}^\infty \frac1{(-T)^{nm}}$. For simplicity, we only consider the two-term model as follows \begin{eqnarray}
f_{1}=&\frac{\alpha H_{0}^{2m}}{(-T)^{m}}+\frac{\gamma H_{0}^{4m}}{(-T)^{2m}}\ ,\label{f1}\\
f_{2}=&\frac{\beta H_{0}^{2m}}{\Omega_{M0} (-T)^{m}}+\frac{\delta H_{0}^{4m}}{\Omega_{M0} (-T)^{2m}}\ ,\label{f2}
\end{eqnarray}
\noindent
where $m$, $\alpha$, $\beta$, $\gamma$ and $\delta$ are the dimensionless parameters to be determined, and $H_{0}$ and $\Omega_{M0}\equiv \frac{\rho_{M0}}{3H_{0}^{2}}$ are the current values of $H$ and the matter density parameter $\Omega_{M}$. We consider five typical cases of the model listed in Table \ref{table case}. Obviously, Model I and Model II in Refs. \cite{Feng2015,Lin2017} will be recovered when $m=1$ in Case 5 and $m=1/2$, $\gamma=0$ in Case 3, respectively.
\begin{table}[!htbp]
	\centering  
	\caption{Five cases of the generalized power-law model.}\label{table case}
	\begin{tabular}{lcccc}  
		\hline
		Case markers &$\quad\alpha\quad$ &$\quad\beta\quad$ &$\quad\gamma\quad$ &$\quad\delta\quad$
		\\ \hline  
		Case 1 &$\neq0$ &$\neq0$ &$\neq0$ &$\neq0$ \\      
		Case 2 &$0$ &$\neq0$ &$\neq0$ &$\neq0$ \\      
		Case 3 &$\neq0$ &$0$ &$\neq0$ &$\neq0$ \\
		Case 4 &$0$ &$0$ &$\neq0$ &$\neq0$ \\
		Case 5 &$0$ &$\neq0$ &$\neq0$ &$0$ \\ \hline
	\end{tabular}
\end{table}\\
\indent
Substituting $T=-6H^{2}$ into the Friedmann equation \eqref{Feq1}, we can rewrite it in the following form:
\begin{eqnarray}
\begin{split}
E^{2}=&\Omega_{M0}a^{-3}+\Omega_{r0}a^{-4}\\
&+\beta(2m+1)6^{-m}E^{-2m}a^{-3}\\
&+\delta(4m+1)6^{-2m}E^{-4m}a^{-3}\\
&+\alpha(2m-1)6^{-m}E^{2-2m}\\
&+\gamma(4m-1)6^{-2m}E^{2-4m}\label{E},
\end{split}
\end{eqnarray}
with the constraint at present time $(a_{0}=1)$
\begin{eqnarray}
\begin{split}
\Omega_{M0}&+\Omega_{r0}+[\alpha(2m-1)+\beta(2m+1)]6^{-m}\\
&+[\gamma(4m-1)+\delta(4m+1)]6^{-2m}=1\label{DE0},
\end{split}
\end{eqnarray}
where $E\equiv\frac{H}{H_{0}}$, $\Omega_{r0}\equiv\frac{\rho_{r0}}{3H_{0}^{2}}$. Eq. \eqref{E} can also be rewritten as
\begin{equation}
E(a)^{2}=\Omega_{M0}a^{-3}+\Omega_{r0}a^{-4}+\Omega_{DE}(a)E(a)^{2},\label{DE}
\end{equation}
where
\begin{eqnarray}
\begin{split}
\Omega_{DE}(a)=&\beta(2m+1)6^{-m}E(a)^{-2m-2}a^{-3}\\
&+\delta(4m+1)6^{-2m}E(a)^{-4m-2}a^{-3}\\
&+\alpha(2m-1)6^{-m}E(a)^{-2m}\\
&+\gamma(4m-1)6^{-2m}E(a)^{-4m},\label{omigeDE}
\end{split}
\end{eqnarray}
and
\begin{eqnarray}
\begin{split}
\Omega_{DE0}&=\beta(2m+1)6^{-m}+\delta(4m+1)6^{-2m}\\
&\ \ \ +\alpha(2m-1)6^{-m}+\gamma(4m-1)6^{-2m}\\
&=1-\Omega_{M0}-\Omega_{r0}.
\end{split}
\end{eqnarray}

By using the data of CMB\cite{Hu1995,Betoule2014,Collaboration2013}, BAO \cite{Betoule2014,Eisenstein1997,Anderson2013a,Padmanabhan2012,Beutler2011} and joint light-curve analysis \cite{Betoule2014,Shafer2015}, we perform cosmological fit for our model, and list the fitting results in Table \ref{fitting}, where we have fixed $\Omega_{r0}=0.00009$ according to Ref. \cite{Feng2015}. The $\Lambda$CDM result is also listed as a reference.
\renewcommand{\baselinestretch}{1.4}
\begin{table*}[!htbp]
	\centering  
	\caption{Best fitting parameters for Cases 1\ \textendash\ 5.}	
	\begin{tabular}{lcccccc}  
		\hline
		Parameters &Case 1 &Case 2 &Case 3 &Case 4 &Case 5 &$\Lambda$CDM
		\\ \hline  
		$m$ &$1.095^{+0.238}_{-0.194}$ &$1.012^{+0.176}_{-0.172}$ &$0.995^{+0.160}_{-3.366}$ &$0.666^{+0.106}_{-0.120}$ &$1.372^{+0.182}_{-0.189}$ &--\\         
		$\alpha$ &$2.697^{+2.196}_{-4.520}$ &-- &$2.492^{+1.590}_{-3.366}$ &-- &-- &--\\          
		$\beta$ &$0.108^{+0.247}_{-0.194}$ &$0.306^{+0.293}_{-0.240}$ &-- &-- &$1.037^{+0.493}_{-0.387}$ &--\\
		$\gamma$ &$1.242^{+1.128}_{-1.589}$ &$5.151^{+8.271}_{-6.359}$ &$1.627^{+1.155}_{-0.985}$ &$3.808^{+1.521}_{-1.843}$ &$11.141^{+16.238}_{-9.865}$ &--  \\
		$\delta$ &$1.101^{+1.831}_{-1.589}$ &$0.784^{+1.311}_{-1.299}$ &$1.124^{+1.359}_{-1.008}$ &$0.344^{+0.450}_{-0.305}$ &-- &--\\ \hline
		$\Omega_{b0}h^{2}$ &$0.0220^{+0.0003}_{-0.0003}$ &$0.0220^{+0.0003}_{-0.0003}$ &$0.0220^{+0.0003}_{-0.0003}$ &$0.0220^{+0.0003}_{-0.0003}$ &$0.0220^{+0.0003}_{-0.0003}$ &$0.0221^{+0.0002}_{-0.0002}$\\
		$\Omega_{m0}$ &$0.254^{+0.011}_{-0.010}$ &$0.255^{+0.010}_{-0.009}$ &$0.254^{+0.010}_{-0.010}$ &$0.255^{+0.010}_{-0.009}$ &$0.255^{+0.009}_{-0.009}$ &$0.257^{+0.009}_{-0.009}$\\
		$H_{0}$ &$69.00^{+1.37}_{-1.40}$ &$68.75^{+1.34}_{-1.23}$ &$69.04^{+1.26}_{-1.20}$ &$68.79^{+1.30}_{-1.22}$ &$68.83^{+1.27}_{-1.27}$ &$68.03^{+0.74}_{-0.74}$\\ \hline
		$\chi^{2}/d.o.f$ &$683.534/735$ &$683.644/736$ &$683.511/736$ &$683.758/737$ &$683.767/737$ &$684.081/739$\\		\hline
	\end{tabular}\label{fitting}
\end{table*}

\renewcommand{\baselinestretch}{}
Naturally, with $T=-6H^{2}$, at current time, we have $f_{1}=\frac{\alpha}{6^{m}}+\frac{\gamma}{6^{2m}}$ and $f_{2}=\frac{\beta}{6^{m}\Omega_{M0}}+\frac{\delta}{6^{2m}\Omega_{M0}}$. This means that the present values of $f_{1}$ and $f_{2}$ only depend on $\Omega_{M0}$ and the dimensionless parameters in the model but not $H_{0}$. Therefore, according to Table \ref{fitting}, the parameters of the model do not need any fine-tuning.

\begin{figure}[!htbp]
	\includegraphics[width=8.6cm,height=5.0cm]{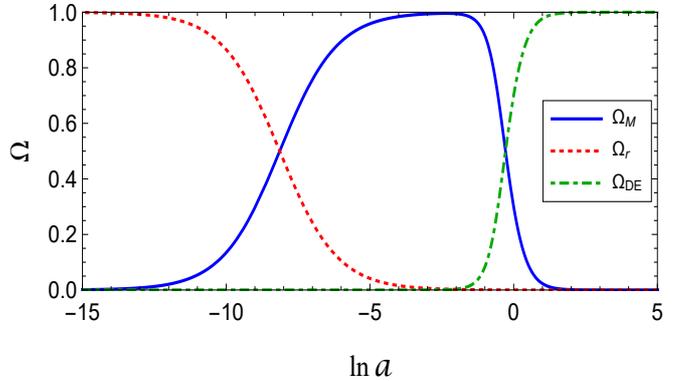}
	\caption{(color online). Evolution of $\Omega_{M}$, $\Omega_{r}$ and $\Omega_{DE}$ with the best-fit values of the parameters for Case 1.}
	\label{fig.Omiga}
\end{figure}

\begin{figure}[!htbp]
	\includegraphics[width=8.6cm,height=6.0cm]{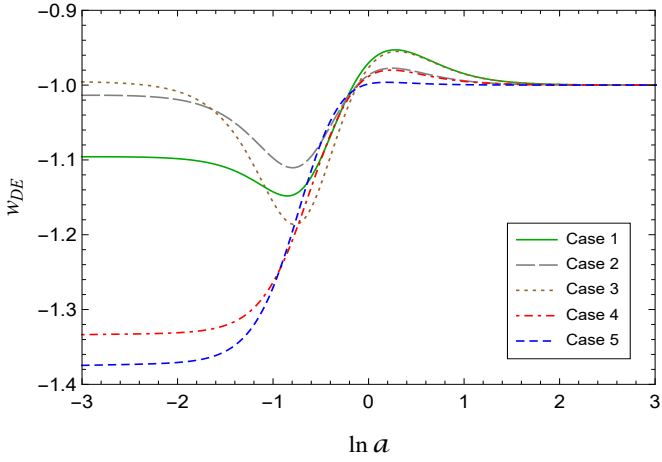}
	\caption{(color online). Evolution of $w_{DE}$ with the best-fit values of the parameters for Cases 1\ \textendash\ 5.}
	\label{fig.wDE}
\end{figure}

With the best-fit values in Table \ref{fitting}, we plot the evolutions of the density parameters for Case 1 in Fig. \ref{fig.Omiga}. Furthermore, from continuity equation, one can obtain the equation of state of effective dark energy $w_{DE}(a)=-1-\frac{1}{3}\frac{d\,ln\,\rho_{DE}}{d\,ln\,a}$. Using Eqs. \eqref{E} and \eqref{omigeDE}, we plot the evolution of $w_{DE}(a)$ for the five cases in Fig. \ref{fig.wDE}. At present time, $w_{DE0}=-0.9704$, $-0.9866$, $-0.9768$, $-0.9883$ and $-0.9985$ for Case 1, 2, 3, 4, and 5, respectively.

In the following sections, the parameters of the cases are set as the best-fit values listed in Table \ref{fitting}.

\section{Dynamical system analysis}\label{secIV}
\subsection{Autonomous system}
In this section, we build the dynamical systems for our model. Defining new variables $X\equiv\Omega_{M}\equiv\frac{\rho_{M}}{3H^{2}}$, $Y\equiv\Omega_{r}\equiv\frac{\rho_{r}}{3H^{2}}$, and $N=ln(a)$, we can rewrite Eqs. \eqref{Feq1}\textendash\eqref{coneq2} as
\begin{eqnarray}
\begin{split}
(1+f_{2}-2Tf_{2}^{'})X+Y-(f_{1}+2Tf_{1}^{'})=1,\label{txyeq}
\end{split}
\end{eqnarray}
\begin{eqnarray}
\begin{split}
\frac{\dot{H}}{H^{2}}=&-[\frac{3}{2}(1+f_{2}-2Tf_{2}^{'})X+2Y]/[1+f_{1}\\
&+5Tf_{1}^{'}+2T^{2}f_{1}^{''}+(Tf_{2}^{'}+2T^{2}f_{2}^{''})X],\label{Heq}
\end{split}
\end{eqnarray}
and
\begin{equation}
\frac{dX}{dN}=\frac{d\Omega_{M}}{dN}=-X(3+2\frac{\dot{H}}{H^{2}}),\label{dseq1}
\end{equation}
\begin{equation}
\frac{dY}{dN}=\frac{d\Omega_{r}}{dN}=-Y(4+2\frac{\dot{H}}{H^{2}}).\label{dseq2}
\end{equation}
For a certain model $f_{1}$ and $f_{2}$, we can get $T=T(X,Y)$ from Eq. \eqref{txyeq}. Then according to Eq. \eqref{Heq}, Eqs. \eqref{dseq1} and \eqref{dseq2} can be written in the forms
\begin{equation}
\frac{dX}{dN}=-X[3-2(\frac{c_{1}X+2Y}{c_{2}+c_{3}X})],\label{dseq1-1}
\end{equation}
\begin{equation}
\frac{dY}{dN}=-Y[4-2(\frac{c_{1}X+2Y}{c_{2}+c_{3}X})],\label{dseq2-1}
\end{equation}
where
\begin{eqnarray}
\begin{split}
&c_{1}=\frac{3}{2}(1+f_{2}-2Tf_{2}^{'}),\\
&c_{2}=1+f_{1}+5Tf_{1}^{'}+2T^{2}f_{1}^{''},\\
&c_{3}=Tf_{2}^{'}+2T^{2}f_{2}^{''}.
\end{split}
\end{eqnarray}
\indent
For our model given by Eqs. \eqref{f1} and \eqref{f2}, we have
\begin{equation}
T=-2^{-1/m}H_{0}^{2}K^{1/m},\label{t(xy)}
\end{equation}
\begin{eqnarray}
\begin{split}
\frac{dX}{dN}=&X\{K^{2}\Omega_{M0}(3X+4Y-3)-6K(m-1)\\
&\times[\beta X(2m+1)+\alpha\Omega_{M0}(2m-1)]\\
&-12(2m-1)[\delta X(4m+1)\\
&+\gamma\Omega_{M0}(4m-1)]\}\\
&/\{\Omega_{M0}[K^{2}+2\alpha K(2m-1)(m-1)\\
&+\gamma (32m^{2}-24m+4)]\\
&+2mX[\beta K(2m+1)+4\delta (4m+1)]\},\label{dseq1-2}
\end{split}
\end{eqnarray}
and
\begin{eqnarray}
\begin{split}
\frac{dY}{dN}=&Y\{K^{2}\Omega_{M0}(3X+4Y-4)-2K\\
&\times[\beta X(8m^{2}-2m-3)+4\alpha\Omega_{M0}(2m^{2}\\
&-3m+1)]-4[\delta X(32m^{2}-4m-3)\\
&+4\gamma\Omega_{M0}(8m^{2}-6m+1)]\}\\
&/\{\Omega_{M0}[K^{2}+2\alpha K(2m-1)(m-1)\\
&+\gamma (32m^{2}-24m+4)]\\
&+2mX[\beta K(2m+1)+4\delta (4m+1)]\},\label{dseq2-2}
\end{split}
\end{eqnarray}
where
\begin{eqnarray}
\begin{split}
K=&\frac{1}{(1-X-Y)\Omega_{M0}}\{\beta X(2m+1)\\
&+\alpha\Omega_{M0}(2m-1)+\{[\beta X(2m+1)\\
&+\alpha\Omega_{M0}(2m-1)]^{2}+4\Omega_{M0}(1-X-Y)\\
&\times[\delta X(4m+1)+\gamma\Omega_{M0}(4m-1)]\}^{\frac{1}{2}}\}.\label{K}
\end{split}
\end{eqnarray}
 Thus Eqs. \eqref{dseq1-2} and \eqref{dseq2-2} are now the 2-dimensional autonomous differential equations with the only two independent variables $X$ and $Y$.

 \subsection{Fixed points and phase space}
 Take Case 1 for example. By making
 \begin{eqnarray}
\frac{dX}{dN}=0,\quad
\frac{dY}{dN}=0,\label{dxy0}
 \end{eqnarray}
  we can find the fixed points of the dynamical system, which are shown in Table \ref{table fix point}. The eigenvalues $(\mu, \nu)$ of the fixed points are also listed in the table.
  Since the fitting results suggest that $m>0$, the fixed points A, B and C are always repeller, saddle and attractor, respectively. Furthermore, we list in Table \ref{table fix point} the behaviors of scale factor $a$ and the deceleration parameter $q$ for each of the fixed points. The fixed points of Cases 2\ \textendash\ 5 have similar properties to Case 1.
\begin{table}[!htbp]
	\centering  
	\caption{The fixed points and their properties for Case 1}\label{table fix point}
	\begin{tabular}{cccccc}  
		\hline
		Fixed\\points \  &$\ (X, Y)\ $ &$\quad(\mu, \nu)\quad$  & \ Stability\quad &$\:\: a(t)\:\: $ &$q$
		\\ \hline  
		A &(0, 1) &$(1, 4m)$  & Repeller &$\propto t^{\frac{1}{2}}$ &1  \\         
		B &(1, 0) &$(-1,3m)$ & Saddle &$\propto t^{\frac{2}{3}}$ &$\frac{1}{2}$   \\        
		C &(0, 0) &$(-4,-3)$ & Attractor &$\propto e^{Ht}$ &-1   \\ \hline
	\end{tabular}
	
	\begin{tabular}{p{0.85\columnwidth}}
		Here $q=-1-\frac{\dot{H}}{H^{2}}$ is the deceleration parameter, $(\mu, \nu)$ are two eigenvalues of each point.
	\end{tabular}			
\end{table}

\begin{figure}[!htbp]
	\includegraphics[width=8.6cm,height=7.5cm]{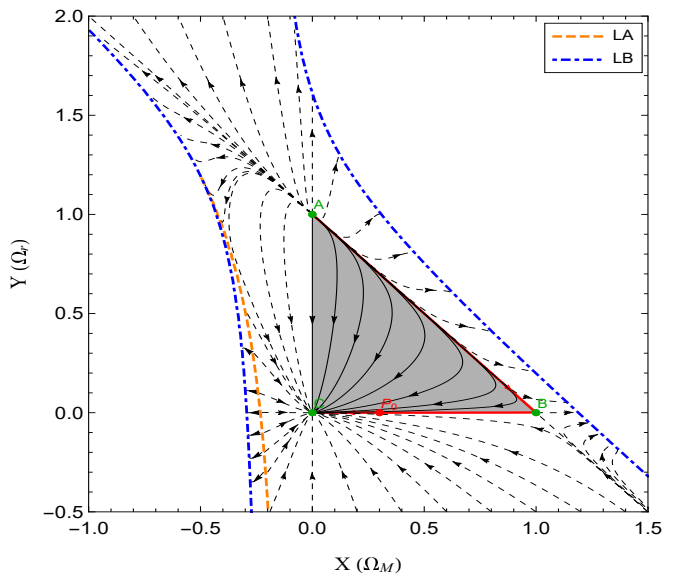}
	\caption{(color online). The phase space for Case 1. The gray area is the physical part of the phase space for which $T=-6H^{2}<0$ and $0\leqslant\Omega_{M0},\Omega_{r}\leqslant1$. The trajectories drawn with solid and dashed lines represent physical and unphysical paths for the system, respectively. The thick solid line is the path which passes through the point $P_{0}(X=0.30021, Y=0.00009)$ representing the current state of the Universe. LA is the line on which Eqs. \eqref{dseq1-2} and \eqref{dseq2-2} become singular. LB is the line on which a common factor of Eqs. \eqref{dseq1-2} and \eqref{dseq2-2} evaluates zero.}\label{fig.dns-I}
\end{figure}

The phase space of Case 1 is illustrated in Fig. \ref{fig.dns-I}. One can see that the evolution of the Universe begins from point A, corresponding to the radiation dominated era, goes near the neighborhood of point B, corresponding to the matter dominated era, then passes through point $P_{0}$, representing the current state of the Universe and finally will arrive at point C, or equivalently $\Omega_{DE}=1$, corresponding to the dark energy dominated era. The phase spaces of Cases 2\ \textendash\ 5 are of the similar behaviors.

\section{Statefinder diagnostic analysis}\label{secV}
Usually, Hubble parameter $H=\frac{\dot{a}}{a}$ and deceleration parameter $q=-1-\frac{\dot{H}}{H}$ are used to describe the expansion of the Universe, but they can not characterize the cosmological models which happen to have the same or very similar current values of $H$ and $q$. To characterize and distinguish our model, in this section, we use statefinder diagnostic  for Cases 1\ \textendash\ 5 and the other models such as
$\Lambda$CDM, the two models in Refs. \cite{Feng2015,Lin2017}, quintessence \cite{Caldwell1997,Scherrer2008,Scherrer2007} and Chaplygin gas \cite{Bilic2001,Kamenshchik2001,Alam2003}. 	
\par
\begin{figure}[!htbp]
	\includegraphics[width=8.6cm,height=5.0cm]{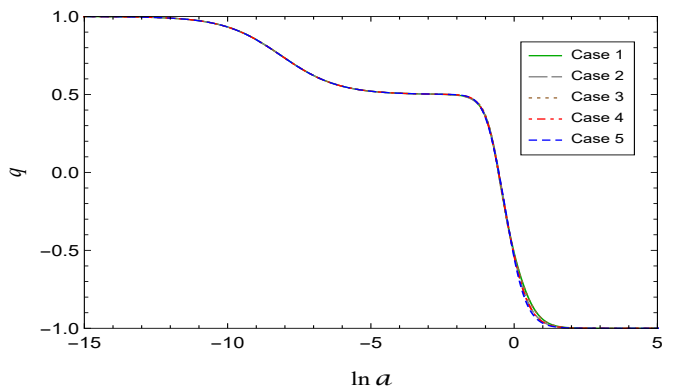}
	\caption{(color online). The evolution of deceleration parameter $q$ for Cases 1\ \textendash\ 5.}\label{q}
\end{figure}

\begin{figure}[!htbp]
	\includegraphics[width=8.6cm,height=5.0cm]{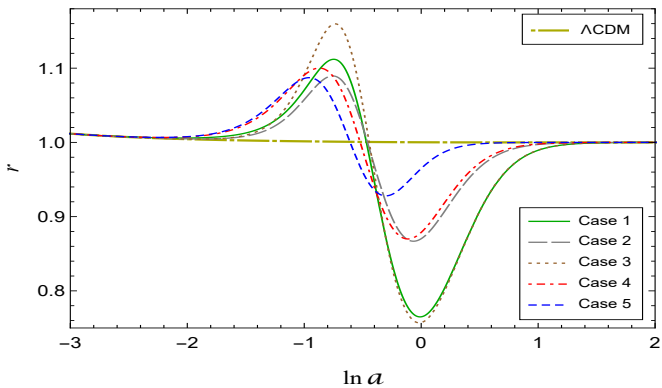}
	\caption{(color online).The evolution of statefinder parameter $r$ for Cases 1\ \textendash\ 5. The horizontal line represents $\Lambda$CDM model.}\label{r}
\end{figure}

\begin{figure}[!htbp]
	\includegraphics[width=8.6cm,height=5.0cm]{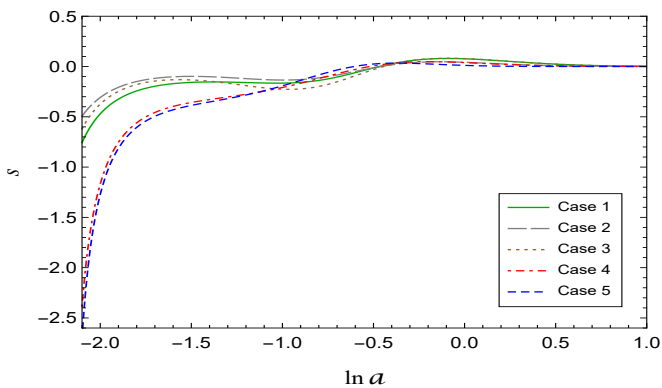}
	\caption{(color online). The evolution of statefinder parameter $s$ for Cases 1\ \textendash\ 5. }\label{s}
\end{figure}

\setlength\parindent{1em}
The statefinder pair $(r,s)$ is defined in Ref. \cite{Sahni2002} as
\begin{eqnarray}
r=\frac{\dddot{a}}{aH^{3}},\label{eq-r}\\
s=\frac{r-1}{3(q-\frac{1}{2})}.\label{eq-s}
\end{eqnarray}
$q$ and $r$ can be rewritten in the following form:
\begin{eqnarray}
q=-1-\frac{1}{H}\frac{dH}{dN},\label{q1}
\end{eqnarray}
and
\begin{eqnarray}
r=1+3\frac{1}{H}\frac{dH}{dN}+(\frac{1}{H}\frac{dH}{dN})^{2}+\frac{1}{H}\frac{d^{2}H}{dN^{2}}.\label{r1}
\end{eqnarray}
Using Eq. \eqref{E}, we plot the evolution of $q$, $r$, $s$ as functions of $N$ numerically in Figs. \ref{q}, \ref{r} and \ref{s}.

From Fig. \ref{q}, one can see that for most part of the evolution, the parameter $q$ is not very distinguishable among the cases. However, in Figs \ref{r} amd \ref{s}, the evolutionary curves of $r$ and $s$ help separate different cases, which admits the significance of the statefinder parameters $r$ and $s$.

\begin{figure}[!htbp]
	\includegraphics[width=8.6cm,height=5.8cm]{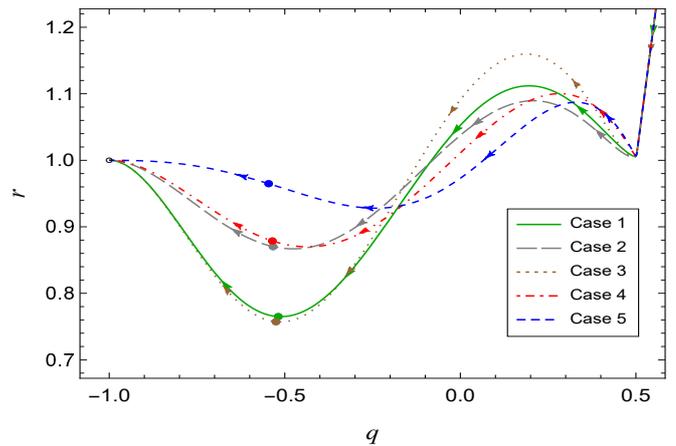}
	\caption{(color online).The $q-r$ diagram of Cases 1\ \textendash\ 5. The filled rounded markers on each line locate the current values of the statefinder pair$ (q,r)$ for the corresponding cases and the empty rounded marker is the point $(q=-1,r=1)$.}\label{q-r}
\end{figure}

\begin{figure}[!htbp]
	\includegraphics[width=8.6cm,height=5.8cm]{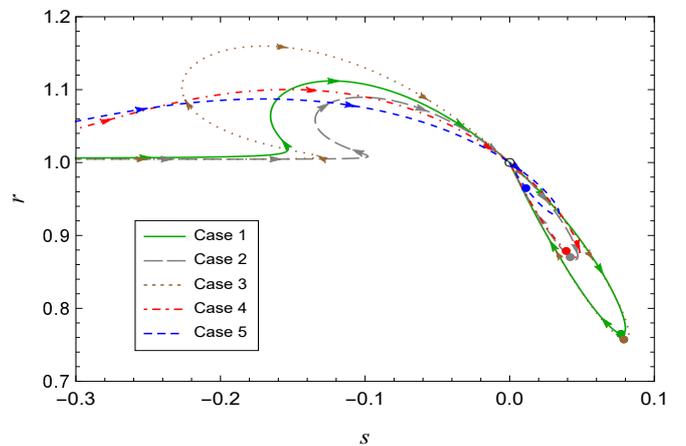}
	\caption{(color online).The $s-r$ diagram of Cases 1\ \textendash\ 5. The filled rounded markers on each line locate the current values of the statefinder pair$ (s,r)$ for the corresponding cases and the empty rounded marker is the point $(s=0,r=1)$.}\label{s-r}
\end{figure}

Figs. \ref{q-r} and \ref{s-r} illustrate the $q-r$ diagram and $s-r$ diagram of the cases. Figs. \ref{r}, \ref{q-r} and \ref{s-r} further demonstrate the merits of the statefinder pair $(r,s)$ and help differentiate the cases. In Fig. \ref{r}, one can observe that for all the cases, the evolutionary curves of our model cross the $\Lambda$CDM line from above to below in the past and will converge to the de Sitter evolution eventually. The loops in Fig. \ref{s-r} also depict this characteristic. For all the cases of our model, the trails of $s-r$ diagram pass through the point $(s=0,r=1)$, which represents the $\Lambda$CDM model. And after some detours, the trails come back to this point. According to Eqs. \eqref{eq-r} and \eqref{eq-s}, the loops suggest that the crossings of $r$ curves from above the $\Lambda$CDM line to below and their convergence to the $\Lambda$CDM model all happen in the $q<1/2$ region.

\begin{figure}[!htbp]
	\includegraphics[width=8.6cm,height=6.0cm]{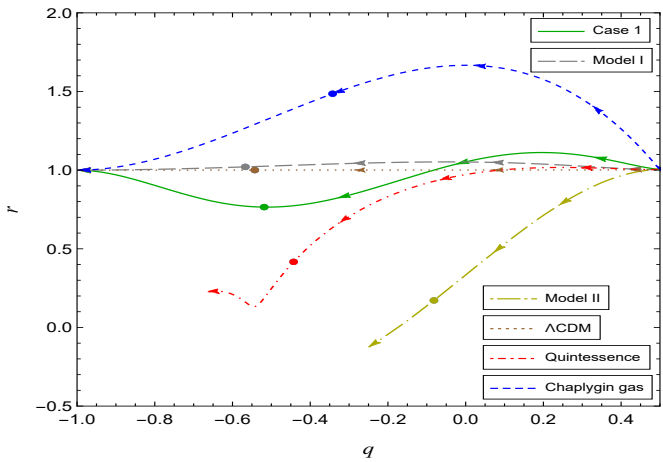}
	\caption{(color online). The $q-r$ diagram of Case 1, Model I and Model II in Refs. \cite{Feng2015,Lin2017}, $\Lambda$CDM, quintessence and Chaplygin gas. The rounded markers on their corresponding lines locate the current values of the statefinder pair$ (q,r)$ for the corresponding models.}\label{q-rw}
\end{figure}

\begin{figure}[!htbp]
	\includegraphics[width=8.6cm,height=6.0cm]{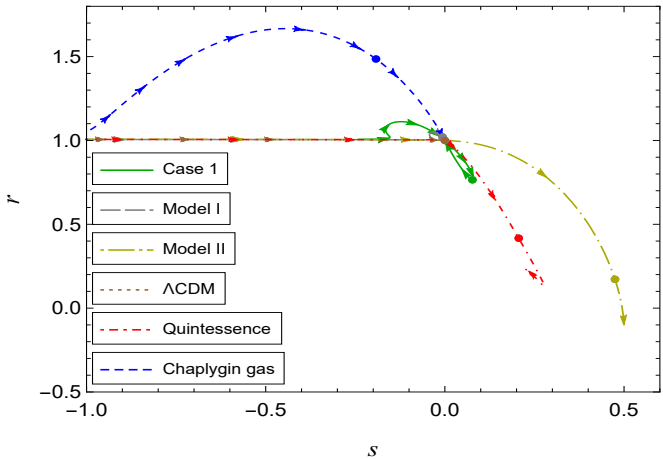}
\caption{(color online). The $s-r$ diagram of Case 1, Model I and Model II in Refs. \cite{Feng2015,Lin2017}, $\Lambda$CDM, quintessence and Chaplygin gas. The rounded markers on their corresponding lines locate the current values of the statefinder pair$ (s,r)$ for the corresponding models.}\label{s-rw}
\end{figure}

Furthermore, using the statefinder parameters, we compare Case 1 of our model with other cosmological models such as the two models in Refs. \cite{Feng2015,Lin2017}, quintessence and Chaplygin gas model, and plot the $q-r$ and $s-r$ diagrams in Figs. \ref{q-rw} and \ref{s-rw}. It can be observed from these two graphs that the cases of our model can not only be told apart from each other, but also be distinguished from other models.

\section{$Om$ dianostic analysis}\label{secVI}
The $Om$ diagnostic \cite{Sahni2008} is another useful tool to resolve the degeneracy amongst dark energy models. It is a combination of the Hubble parameters and the cosmological redshift, and provides a null test of dark energy being a cosmological constant $\Lambda$. $Om$ is determined by
\begin{eqnarray}
Om(z)\equiv\frac{E^{2}(z)-1}{(1+z)^{3}-1}
\end{eqnarray}
as a function of the redshift $z$, where $E(z)=\frac{H(z)}{H_{0}}$, and for the model at hand, it is given by Eq. \eqref{E}.

For $\Lambda$CDM model, since $E^{2}(z)=\Omega_{M0}(1+z)^{3}+(1-\Omega_{M0})$, $Om(z)$ is a horizontal line in $z-Om$ diagram, namely, $Om(z)=\Omega_{M0}$, while dynamical dark energy models give curves. For quintessence dark energy model where $w_{DE}>-1$, $Om(z)$ has a negative slope and $Om(z)>\Omega_{M0}$, while for phantom model where $w_{DE}<-1$, it has a positive slope and $Om(z)<\Omega_{M0}$.

\begin{figure}[!htbp]
	\includegraphics[width=8.6cm,height=6.5cm]{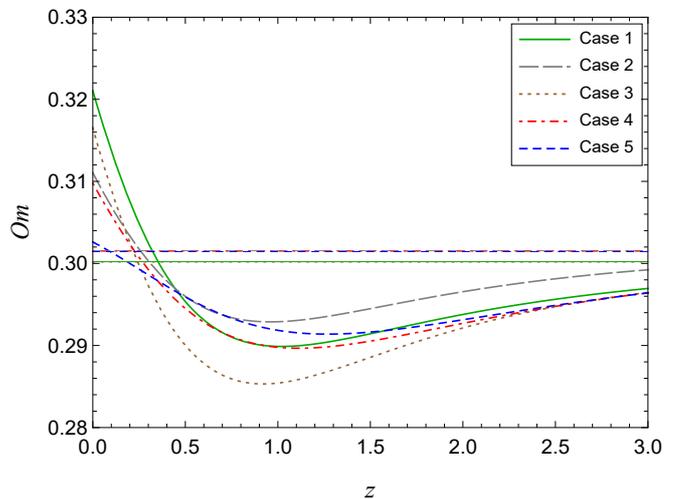}
	\caption{(color online). The $Om$ diagrams of Cases 1\ \textendash\ 5 with the best-fit $\Omega_{M0}$ as 0.300, 0.302, 0.300, 0.301, 0.301, respectively. The horizontal lines represent  $\Lambda$CDM models with the same values of best-fit $\Omega_{M0}$ as the corresponding cases.}\label{Om1}
\end{figure}

\begin{figure}[!htbp]
	\includegraphics[width=8.6cm,height=6.5cm]{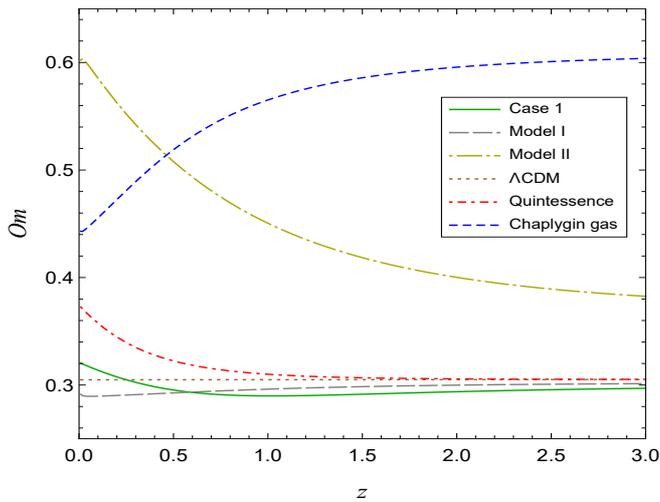}
	\caption{(color online). The $Om$ diagrams of Case 1, Model I, Model II, $\Lambda$CDM, quintessence and Chaplygin gas. }\label{Om2}
\end{figure}

For the five cases of our model we plot $Om(z)$ in Fig. \ref{Om1}. From the illustration one can see that in the past the curves of all the cases have crossed the corresponding horizontal lines representing $\Lambda$CDM models with the same values of best-fit $\Omega_{M0}$ as the corresponding cases, which is in accordance with the evolution of $w_{DE}$ plotted in Fig. \ref{fig.wDE} and statefinder $r$ plotted in Fig. \ref{r}.  We also plot the $Om$ diagram of other dark energy models in Fig. \ref{Om2} for contrast.

It has been studied that various types of singularity might occur in phantom scenarios \cite{Nojiri2005}. Some mechanisms have been proposed to avoid these catastrophic singularities. For example, it has been shown that these singularities in the future of the cosmic evolution might be avoided if we take suitable potential term \cite{Hao2003,Liu2003,Hao2004}, the phantom-generalized Chaplygin gas \cite{Zhai2006,Hao2005}, and the torsion cosmology \cite{Li:2009zzc}. By using the analyses above, we found that the evolution of the equation of state parameter $w$ with cosmic time can cross the divide $w=-1$ from below to above and approach to $w=-1$ in the future, that is to say, all kinds of generalized power-law torsion-matter coupling $f(T)$ model are able to avoid the catastrophic big or little rip \cite{Xi2012}.


\section{Concluding remarks}\label{secVII}
The two non-minimal torsion-matter coupling $f(T)$ models
established in the previous papers \cite{Feng2015,Lin2017}
are highly successful in describing the observations of the
expansion history of the Universe and its large scale structure,
as well as the Solar system tests of gravity.
We have generalized those two models in this paper,
and investigated several different cases of
this generalized power-law
torsion-matter coupling $f(T)$ model.

Using the dynamical system analysis,
we find that the generalized power-law torsion-matter
coupling $f(T)$ model confirms the usual physical area of the phase space.
The stability analysis of the critical points shows that
the model can reproduce the different expansion phases of the Universe,
i.e. the radiation dominated era,
the matter dominated era and the dark energy dominated era, and it has a de Sitter expansion fate in the future.

Employing the statefinder diagnostic, we find that the five cases of our model can be distinguished from each other, and from other cosmology models such as the two models in Refs. \cite{Feng2015,Lin2017}, $\Lambda$CDM, quintessence and Chaplygin gas. The evolutionary trajectories of the statefinder of the cases also have common properties in that they all pass the $r=1$ state from above to below in the past and that they will converge to this state eventually in the future.
	
Moreover, by using the $Om$ diagnostic, we confirm that the five cases of our model all cross $w=-1$ divide, which is in accordance with the evolution of $w_{DE}$ and statefinder diagnostic. For Cases 1, 2, 3, 4 and 5, the crossings happen at $z=$ 0.358, 0.311, 0.245, 0.274 and 0.200, respectively. It is worth noting that crossing $w=-1$ divide happened lately and the decrease of the energy density resulting from the crossing of $w$ will make the catastrophic fate avoided.

\section*{Acknowledgments}
This work is supported by the National Science Foundation of China grant No. 10671128, and the Key Project of Chinese Ministry of Education grant No. 211059.

\bibliographystyle{spphys}
\bibliography{myreference}

\begin{thebibliography}{10}
\providecommand{\url}[1]{{#1}}
\providecommand{\urlprefix}{URL }
\expandafter\ifx\csname urlstyle\endcsname\relax
  \providecommand{\doi}[1]{DOI \discretionary{}{}{}#1}\else
  \providecommand{\doi}{DOI \discretionary{}{}{}\begingroup
  \urlstyle{rm}\Url}\fi

\bibitem{Sotiriou:2008rp}
T.P. Sotiriou, V.~Faraoni, Rev. Mod. Phys. \textbf{82}, 451 (2010)

\bibitem{DeFelice:2010aj}
A.~De~Felice, S.~Tsujikawa, Living Rev. Rel. \textbf{13}, 3 (2010)

\bibitem{Zhang:2011uf}
H.~Zhang, X.Z. Li, Phys. Lett. B \textbf{715}, 15 (2012)

\bibitem{Li:2009zzc}
X.Z. Li, C.B. Sun, P.~Xi, Phys. Rev. D \textbf{79}, 027301 (2009)

\bibitem{Ao:2010mg}
X.C. Ao, X.Z. Li, P.~Xi, Phys. Lett. B \textbf{694}, 186 (2011)

\bibitem{Ao:2011kc}
X.C. Ao, X.Z. Li, JCAP \textbf{1202}, 003 (2012)

\bibitem{Einstein:1928}
A.Einstein, Sitz. Preuss. Akad. Wiss. \textbf{p. 217; ibidp. 224} (1928)

\bibitem{Ferraro2006}
R.~Ferraro, F.~Fiorini, Phys. Rev. D \textbf{75}, 084031 (2007)

\bibitem{Linder2010}
E.V. Linder, Phys. Rev. D \textbf{81}, 127301 (2010).
\newblock [Erratum: Phys. Rev.D82,109902(2010)]

\bibitem{Aly:2015fda}
A.A. Aly, M.M. Selim, Eur. Phys. J. Plus \textbf{130}(8), 164 (2015)

\bibitem{Cai2015}
Y.F. Cai, S.~Capozziello, M.~De~Laurentis, E.N. Saridakis, Rept. Prog. Phys.
  \textbf{79}(10), 106901 (2016)

\bibitem{Farrugia:2016}
G.~Farrugia, J.L. Said, M.L. Ruggiero, Phys. Rev. D \textbf{93}(10), 104034
  (2016)

\bibitem{Nunes:2016plz}
R.C. Nunes, A.~Bonilla, S.~Pan, E.N. Saridakis, Eur. Phys. J. C \textbf{77}(4),
  230 (2017)

\bibitem{Oikonomou:2016jjh}
V.K. Oikonomou, E.N. Saridakis, Phys. Rev. D \textbf{94}(12), 124005 (2016)

\bibitem{Lin2016}
R.H. Lin, X.H. Zhai, X.Z. Li, JCAP \textbf{1703}(03), 040 (2017)

\bibitem{Hohmann2017}
M.~Hohmann, L.~Jarv, U.~Ualikhanova, Phys. Rev. D \textbf{96}, 043508 (2017)

\bibitem{Qi:2017xzl}
J.Z. Qi, S.~Cao, M.~Biesiada, X.~Zheng, H.~Zhu, Eur. Phys. J. C \textbf{77}(8),
  502 (2017)

\bibitem{Sk:2017ucb}
N.~Sk, Phys. Lett. B \textbf{775}, 100 (2017)

\bibitem{Capozziello:2017bxm}
S.~Capozziello, G.~Lambiase, E.N. Saridakis, Eur. Phys. J. C \textbf{77}(9),
  576 (2017)

\bibitem{Oikonomou:2017isf}
V.K. Oikonomou, Phys. Rev. D \textbf{95}(8), 084023 (2017)

\bibitem{Pace:2017aon}
M.~Pace, J.L. Said, Eur. Phys. J. C \textbf{77}(2), 62 (2017)

\bibitem{Zhai:2017yqd}
X.H. Zhai, R.H. Lin, C.J. Feng, X.Z. Li, Phys. Rev. D \textbf{95}(10), 104030
  (2017)

\bibitem{Harko:2014aja}
T.~Harko, F.S.N. Lobo, G.~Otalora, E.N. Saridakis, JCAP \textbf{1412}, 021
  (2014)

\bibitem{Kofinas:2014owa}
G.~Kofinas, E.N. Saridakis, Phys. Rev. D \textbf{90}, 084044 (2014)

\bibitem{Bahamonde2016}
S.~Bahamonde, C.G. B{\"o}hmer, Eur. Phys. J. C \textbf{76}(10), 578 (2016)

\bibitem{Capozziello2016}
S.~Capozziello, M.~De~Laurentis, K.F. Dialektopoulos, Eur. Phys. J. C
  \textbf{76}(11), 629 (2016)

\bibitem{Jawad2015}
A.~Jawad, Eur. Phys. J. Plus \textbf{130}(5), 94 (2015)

\bibitem{Kofinas:2014daa}
G.~Kofinas, E.N. Saridakis, Phys. Rev. D \textbf{90}, 084045 (2014)

\bibitem{Harko2014}
T.~Harko, F.S.N. Lobo, G.~Otalora, E.N. Saridakis, Phys. Rev. D \textbf{89},
  124036 (2014)

\bibitem{Carloni2015}
S.~Carloni, F.S.N. Lobo, G.~Otalora, E.N. Saridakis, Phys. Rev. D \textbf{93},
  024034 (2016)

\bibitem{Feng2015}
C.J. Feng, F.F. Ge, X.Z. Li, R.H. Lin, X.H. Zhai, Phys. Rev. D \textbf{92}(10),
  104038 (2015)

\bibitem{Lin2017}
R.H. Lin, X.H. Zhai, X.Z. Li, Eur. Phys. J. C \textbf{77}(8), 504 (2017)

\bibitem{Xi2012}
P.~Xi, X.H. Zhai, X.Z. Li, Phys. Lett. B \textbf{706}, 482 (2012)

\bibitem{Tamanini:2012hg}
N.~Tamanini, C.G. B\"ohmer, Phys. Rev. D \textbf{86}, 044009 (2012)

\bibitem{Hervik}
{\O}.~Gr{\o}n, S.~Hervik, \emph{Einstein’s general theory of relativity:with
  modern applications in cosmology} (Springer Science \& Business Media, 2007)

\bibitem{Bertolami2008}
O.~Bertolami, F.S.N. Lobo, J.~Paramos, Phys. Rev. D \textbf{78}, 064036 (2008)

\bibitem{Hu1995}
W.~Hu, N.~Sugiyama, Astrophys. J. \textbf{471}, 542 (1996)

\bibitem{Betoule2014}
M.~Betoule, et~al., Astron. Astrophys. \textbf{568}, A22 (2014)

\bibitem{Collaboration2013}
P.~Collaboration, P.A.R. Ade, N.~Aghanim, et~al., Astron. Astrophys.
  \textbf{571}, A16 (2014)

\bibitem{Eisenstein1997}
D.J. Eisenstein, W.~Hu, Astrophys. J. \textbf{496}, 605 (1998)

\bibitem{Anderson2013a}
L.~Anderson, E.~Aubourg, S.~Bailey, et~al., Mon. Not. Roy. Astron. Soc.
  \textbf{441}(1), 24 (2014)

\bibitem{Padmanabhan2012}
N.~Padmanabhan, X.~Xu, D.J. Eisenstein, et~al., Mon. Not. Roy. Astron. Soc.
  \textbf{427}(3), 2132 (2012)

\bibitem{Beutler2011}
F.~Beutler, C.~Blake, M.~Colless, et~al., Mon. Not. Roy. Astron. Soc.
  \textbf{416}, 3017 (2011)

\bibitem{Shafer2015}
D.L. Shafer, Phys. Rev. D \textbf{91}(10), 103516 (2015)

\bibitem{Caldwell1997}
R.R. Caldwell, R.~Dave, P.J. Steinhardt, Phys. Rev. Lett. \textbf{80}, 1582
  (1998)

\bibitem{Scherrer2008}
R.J. Scherrer, A.A. Sen, Phys. Rev. D \textbf{78}, 067303 (2008)

\bibitem{Scherrer2007}
R.J. Scherrer, A.A. Sen, Phys. Rev. D \textbf{77}, 083515 (2008)

\bibitem{Bilic2001}
N.~Bilic, G.B. Tupper, R.D. Viollier, Phys. Lett. B \textbf{535}, 17 (2002)

\bibitem{Kamenshchik2001}
A.{\relax Yu}. Kamenshchik, U.~Moschella, V.~Pasquier, Phys. Lett. B
  \textbf{511}, 265 (2001)

\bibitem{Alam2003}
U.~Alam, V.~Sahni, T.D. Saini, A.A. Starobinsky, Mon. Not. Roy. Astron. Soc.
  \textbf{344}, 1057 (2003)

\bibitem{Sahni2002}
V.~Sahni, T.D. Saini, A.A. Starobinsky, U.~Alam, JETP Lett. \textbf{77}, 201
  (2003).
\newblock [Pisma Zh. Eksp. Teor. Fiz.77,249(2003)]

\bibitem{Sahni2008}
V.~Sahni, A.~Shafieloo, A.A. Starobinsky, Phys. Rev. D \textbf{78}, 103502
  (2008)

\bibitem{Nojiri2005}
S.~Nojiri, S.D. Odintsov, S.~Tsujikawa, Phys. Rev. D \textbf{71}, 063004 (2005)

\bibitem{Hao2003}
J.G. Hao, X.Z. Li, Phys. Rev. D \textbf{67}, 107303 (2003)

\bibitem{Liu2003}
D.J. Liu, X.Z. Li, Phys. Rev. D \textbf{68}, 067301 (2003)

\bibitem{Hao2004}
J.G. Hao, X.Z. Li, Phys. Rev. D \textbf{70}, 043529 (2004)

\bibitem{Zhai2006}
X.H. Zhai, Y.D. Xu, X.Z. Li, Int. J. Mod. Phys. D \textbf{15}, 1151 (2006)

\bibitem{Hao2005}
J.G. Hao, X.Z. Li, Phys. Lett. B \textbf{606}, 7 (2005)

\end{thebibliography}
\end{document}